\documentclass[12pt]{article}
\usepackage[a4paper]{geometry}
\usepackage[italian,english]{babel}
\usepackage[T1]{fontenc}

\usepackage{amsmath, accents}
\usepackage{amsfonts}
\usepackage{amstext}
\usepackage{amssymb}
\usepackage{amsthm}
\usepackage{amscd}
\usepackage{mathrsfs}
\usepackage{bbold}
\usepackage{dsfont}
\usepackage{bbm}

\usepackage[pagebackref,draft=false]{hyperref}
\hypersetup{colorlinks,
linkcolor=myrefcolor,
citecolor=mycitecolor,
urlcolor=myurlcolor}

\usepackage[capitalize]{cleveref}
\usepackage{cite}
\usepackage{caption}
\usepackage{etaremune}

\usepackage{xcolor}
\definecolor{myurlcolor}{rgb}{0,0,0.4}
\definecolor{mycitecolor}{rgb}{0,0.5,0}
\definecolor{myrefcolor}{rgb}{0.5,0,0}
\usepackage{graphicx}
\usepackage{tikz}
\usepackage{tikz-cd}
\usetikzlibrary{graphs,decorations.pathmorphing,decorations.markings}

\usepackage{lipsum}
\usepackage{etoolbox}
\usepackage{makeidx}
\usepackage{sectsty}
\usepackage{dsfont}
\usepackage{enumitem} 
\usepackage[]{latexsym}
\usepackage{braket}
\usepackage{caption}
\usepackage[utf8]{inputenx}
\usepackage{lmodern}
\usepackage{textcomp}
\usepackage{microtype}
\usepackage{totcount}
\usepackage{blindtext}

\newtheorem{theorem}{Theorem}[section]
\newtheorem{remark}[theorem]{Remark}

\newtheorem{proposition}[theorem]{Proposition}

\newtheorem*{proof*}{Proof}

\newcommand{\be}{\begin{equation}}
\newcommand{\ee}{\end{equation}}
\newcommand{\bea}{\begin{eqnarray}}
\newcommand{\eea}{\end{eqnarray}}

\newcommand{\ac}{\mathscr{S}}

\newcommand{\m}{\mathscr{M}}

\newcommand{\pe}{\mathcal{P}(\mathbb{E})}

\newcommand{\fpe}{\mathcal{F}_{\mathcal{P}(\mathbb{E})}}

\newcommand{\pssos}{\Pi^\star_\Sigma \Omega^\Sigma}
\newcommand{\os}{\Omega^\Sigma}

\newcommand{\elag}{\mathcal{E}\mathscr{L}}

\newcommand{\dd}{{\rm d}}
\newcommand{\de}{\partial}

\title{The Geometry of the solution space of first order Hamiltonian field theories III: Palatini's formulation of General Relativity}

\author{F. M. Ciaglia$^{2,6}$ \href{https://orcid.org/0000-0002-8987-1181}{\includegraphics[scale=0.7]{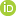}}, F. Di Cosmo$^{1,2,7}$ \href{https://orcid.org/0000-0003-0256-5913}{\includegraphics[scale=0.7]{ORCID.png}}, A. Ibort$^{1,2,8}$ \href{https://orcid.org/0000-0002-0580-5858}{\includegraphics[scale=0.7]{ORCID.png}}, \\ G. Marmo$^{3,4,9}$ \href{https://orcid.org/0000-0003-2662-2193}{\includegraphics[scale=0.7]{ORCID.png}}, L. Schiavone$^{3,5,10}$  \href{https://orcid.org/0000-0002-1817-5752}{\includegraphics[scale=0.7]{ORCID.png}}, A. Zampini$^{3,5,11}$ \href{https://orcid.org/0000-0003-0980-6003}{\includegraphics[scale=0.7]{ORCID.png}} \\
\footnotesize{$^{1}$\textit{ICMAT, Instituto de Ciencias Matem\'{a}ticas (CSIC-UAM-UC3M-UCM)}} \\
\footnotesize{$^{2}$\textit{Departamento de Matem\'aticas, Universidad Carlos III de Madrid, Legan\'es, Madrid, Spain}} \\
\footnotesize{$^{3}$\textit{INFN-Sezione di Napoli, Naples, Italy}} \\
\footnotesize{$^{4}$\textit{Dipartimento di Fisica ``E. Pancini'', Universit\`a di Napoli Federico II,  Naples, Italy}} \\
\footnotesize{$^{5}$\textit{Dipartimento di Matematica e Applicazioni "Renato Caccioppoli", Università di Napoli Federico II, Napoli, Italy}} \\
\footnotesize{$^{6}$\textit{e-mail: \texttt{fciaglia[at]math.uc3m.es
}}} \,\, 
\footnotesize{$^{7}$\textit{e-mail: \texttt{fcosmo[at]math.uc3m.es}}} \\
\footnotesize{$^{8}$\textit{e-mail: \texttt{albertoi[at]math.uc3m.es}}} \,\,  
\footnotesize{$^{9}$\textit{e-mail: \texttt{marmo[at]na.infn.it}}} \\ 
\footnotesize{$^{10}$\textit{e-mail: \texttt{luca.schiavone[at]unina.it}}} \,\,  
\footnotesize{$^{11}$\textit{e-mail: \texttt{azampini[at]na.infn.it}}} 
}

\begin{document}

\maketitle

\tableofcontents

\section*{Introduction}
\label{Sec:Intro}
\addcontentsline{toc}{section}{\nameref{Sec:Intro}}

In the first two parts of this series of papers \cite{Ciaglia-DC-Ibort-Marmo-Schiav-Zamp2021-Cov_brackets_toappear, Ciaglia-DC-Ibort-Mar-Schiav-Zamp2022-Non_abelian} and in some preceding works \cite{Ciaglia-DC-Ibort-Marmo-Schiav2020-Jacobi_Particles, Ciaglia-DC-Ibort-Marmo-Schiav2020-Jacobi_Fields} the authors dealt with the problem of equipping the space of solutions of the equations of motion of a large class of field theories, among which gauge theories, with a Poisson structure.
In this last manuscript we are going to consider the same problem within a relevant case of field theory not covered by the previous contributions, namely General Relativity.
Indeed, it is the unique (classical) field theory describing a fundamental interaction which was not considered yet in \cite{Ciaglia-DC-Ibort-Marmo-Schiav-Zamp2021-Cov_brackets_toappear, Ciaglia-DC-Ibort-Mar-Schiav-Zamp2022-Non_abelian}.

In the first two contributions of this series, we showed that a Poisson bracket on the space of solutions can be given by using a pre-symplectic $2$-form which naturally emerges within the multisymplectic formulation of field theories.
The aim of this paper is to construct such a pre-symplectic structure and its related bracket within General Relativity.
As we will see, the multisymplectic formulation of General Relativity presents some peculiar difficulties with respect to gauge theories dealt with in \cite{Ciaglia-DC-Ibort-Marmo-Schiav-Zamp2021-Cov_brackets_toappear, Ciaglia-DC-Ibort-Mar-Schiav-Zamp2022-Non_abelian}.
Indeed, as we explain in Sec. \ref{Subsec:Variational formulation of General Relativity: from Einstein-Hilbert to Palatini}, \ref{Subsec:The topological limit and the Palatini's constraint} and \ref{Subsec:Schwinger-Weiss variational principle and the formalism near a slice} General Relativity in the Palatini's approach can be formulated as a Yang-Mills theory in a particular limit that we call \textit{topological limit} and by considering an external constraint that we call \textit{Palatini's constraint}.
For this reason in order to deal with General Relativity, we will adapt the formalism used in \cite{Ciaglia-DC-Ibort-Marmo-Schiav-Zamp2021-Cov_brackets_toappear, Ciaglia-DC-Ibort-Mar-Schiav-Zamp2022-Non_abelian} to such a "constrained situation" in Sec. \ref{Sec:Multisymplectic formulation of first order Hamiltonian Field Theories and Lagrange multipliers} where we show how to implement the use of Lagrange multipliers within the multisymplectic formalism. 
With this machinery in hand, we argue, in Sec. \ref{Subsec:Covariant brackets and DeWitt's formula}, how to construct a Poisson bracket on the space of solutions of Einstein's equations in vacuum.

Sec. \ref{Subsec:Covariant brackets and DeWitt's formula} also ends with a conjecture that we explain below.
Without recalling all the motivations behind the problem of equipping the space of solutions with a Poisson structure, for which we refer to the opening paper of this series \cite{Ciaglia-DC-Ibort-Marmo-Schiav-Zamp2021-Cov_brackets_toappear}, let us just mention that such a problem was first considered by \textit{R. E. Peierls} in a paper of $1952$ \cite{Peierls1952-Commutation_laws}.
Very briefly, in that paper \textit{Peierls} gave an algorithm to construct a Poisson bracket between smooth functions on the space of extrema of an action functional.
In particular, given an action functional $\ac$ describing a field theory over $\mathbb{R}^4$, and two smooth real-valued functions on the space of its extrema, say $F$ and $G$, the \textit{Peierls' bracket} between $F$ and $G$ is given by:
\be
\left\{\, F,\, G \,\right\}_{\phi} \,=\, \int_{\mathbb{R}^4 \times \mathbb{R}^4} \frac{\delta F}{\delta \phi} (x) \mathscr{G}^{(c)}(x,\, y) \frac{\delta F}{\delta \phi} (y) \dd^4 y \dd^4 x \,, 
\ee
where $\phi$ are fields of the theory being extrema of $\ac$, $x$ and $y$ are points in $\mathbb{R}^4$ and $\mathscr{G}^{(c)}$ is the causal Green's function providing a solution of the linearization of the equations of motion of the theory.
As it was showed by \textit{B. DeWitt} \cite{DeWitt1965-Groups_Fields, DeWitt1960-Commutators_Quant_Grav}, when the field theory under investigation exhibits gauge symmetries, Peierls' bracket is given by the expression above, provided the fields $\phi$ obey some additional conditions.
In particular, within General Relativity, he proved that such supplementary conditions are \cite{DeWitt1960-Commutators_Quant_Grav}:
\be \label{Eq:DeWitt condition}
\left(\, g^{\mu \nu} g^{\rho \sigma} - \frac{1}{2} g^{\mu \rho} g^{\nu \sigma} \, \right) \, \nabla_\nu \delta g_{\rho \sigma} \,=\, 0 \,,
\ee
where $g$ are the fields of the theory (i.e. the metric of the space-time), $\nabla$ is the (Levi-Civita) covariant derivative and $\delta g$ is a solution of the linearized Einstein's equations.

\noindent In the first two contributions of this series we showed that when gauge symmetries are present, the possibility of defining a Poisson bracket on the space of solutions is related with the existence of a particular geometric structure, namely a flat connection $P$ on a particular bundle related with the theory.
When such a flat connection exists, we proved that the Poisson bracket is given by:
\be \label{Eq:covariant bracket}
\left\{\, F,\, G \,\right\} \,=\, X_G(F) \,,
\ee
where $X_G$ is the Hamiltonian vector field associated to $G$ via a canonical pre-symplectic structure on the space of solutions, provided with the additional condition:
\be \label{Eq:horizontality}
P(X_G) \,=\, 0 \,
\ee
saying that $X_G$ is horizontal with respect to the connection $P$.

\noindent On the other hand, when such a connection can not be chosen to be flat we also argued that it seems that if one wants to use only canonical constructions, one is forced to define the Poisson structure on an enlarged space which, by the way, has a physical interpretation, at least for non-Abelian gauge theories.

\noindent With all this in mind, the conjecture with whom we end Sec. \ref{Subsec:Covariant brackets and DeWitt's formula} is that the supplementary conditions given by \textit{DeWitt} within General Relativity can be interpreted as the condition \eqref{Eq:horizontality} for some flat connection.


\section{Multisymplectic formulation of first order Hamiltonian Field Theories and Lagrange multipliers}
\label{Sec:Multisymplectic formulation of first order Hamiltonian Field Theories and Lagrange multipliers}

We refer to \cite{Ciaglia-DC-Ibort-Marmo-Schiav-Zamp2021-Cov_brackets_toappear, Ciaglia-DC-Ibort-Mar-Schiav-Zamp2022-Non_abelian, Ciaglia-DC-Ibort-Marmo-Schiav-Zamp-2022-Symmetry} and references therein for a complete account on the multisymplectic formulation of first order Hamiltonian field theories and here we limit ourselves to recall the main aspects of the theory in order to recall the notations used in  \cite{Ciaglia-DC-Ibort-Marmo-Schiav-Zamp2021-Cov_brackets_toappear, Ciaglia-DC-Ibort-Mar-Schiav-Zamp2022-Non_abelian} and that we will use in the sequel.

We saw that first order Hamiltonian field theories can be geometrically described as field theories over a space-time $\mathscr{M}$ whose configuration fields can be modelled as sections of a bundle $\pi \;\;:\;\; \mathbb{E} \to \m$, denoted by $\phi$.
The space of configuration fields is a suitable Banach completion of such a space of sections \cite{Ciaglia-DC-Ibort-Marmo-Schiav-Zamp2021-Cov_brackets_toappear, Ciaglia-DC-Ibort-Mar-Schiav-Zamp2022-Non_abelian}, that we denote by $\mathcal{F}_{\mathbb{E}}$.
The carrier space where we settle the Hamiltonian formulation of the theory is the so called Covariant Phase Space, which is denoted by $\pe$ and is a fibre bundle both over $\mathbb{E}$ and over $\m$, whose projections are denoted by $\delta^1_0$ and $\delta_1$ respectively.
The space of dynamical fields of the theory is (a suitable Banach completion of) the space of pairs of sections of $\delta^1_0$ and $\pi$, that we denote by $\fpe$.
Elements of $\fpe$ are denoted by $\chi \,=\, (\phi,\, P)$.
The main points that have to be recalled about the theory developed in \cite{Ciaglia-DC-Ibort-Marmo-Schiav-Zamp2021-Cov_brackets_toappear, Ciaglia-DC-Ibort-Mar-Schiav-Zamp2022-Non_abelian} are the following:
\begin{itemize}
    \item Provided a Hamiltonian, i.e. a local real-valued function $H$ on $\pe$, is fixed, an action functional $\ac$ on $\fpe$ can be defined whose extrema define the space of solutions of the theory, $\elag$, often referred to as solution space.
    \item The solution space $\elag$ is naturally equipped with a canonical differential $2$-form of the type $\pssos$, where $\Sigma$ is any slice of $\m$\footnote{Recall that a slice of $\m$ is a $1$-codimension hypersurface of $\m$ splitting $\m$ into two space-times.}, $\Pi_\Sigma$ is the restriction map $\chi \mapsto \chi\bigr|_\Sigma \in \fpe^\Sigma$, $\fpe^\Sigma$ denotes the space of restrictions of elements of $\fpe$ to the hypersurface $\Sigma$ and $\os$ is a differential $2$-form on $\fpe^\Sigma$.
    The form $\pssos$ is an exact $2$-form being (minus) the differential of a $1$-form of the type $\Pi^\star_\Sigma \alpha^\Sigma$ that naturally emerges from the Schwinger-Weiss variational principle applied to $\ac$, where $\alpha^\Sigma$ is a $1$-form on $\fpe^\Sigma$.
    \item Noting that, at least locally close to any $\Sigma$, the space of dynamical fields is isomorphic to the space of curves over $\fpe^\Sigma$, the isomorphism being denoted by $\varpi$, one is able to prove that (locally, close to $\Sigma$) the space of extrema of $\tilde{\ac}^\epsilon \,:=\, \left(\,\varpi^{-1}\,\right)^\star \ac^\epsilon$\footnote{Where $\ac^\epsilon$ denotes the action functional evaluated on the restrictions of elements of $\fpe$ to a region close to $\Sigma$.}, is isomorphic to the space of solutions of the pre-symplectic Hamiltonian system $\left(\, \fpe^\Sigma,\, \Omega^\Sigma,\, \mathcal{H} \,\right)$, where $\mathcal{H}$ is a suitable Hamiltonian functional associated to $\ac$.
    With this in mind, one proves that $\pssos$ is symplectic or just pre-symplectic depending on whether the differential form $\Omega^\Sigma_\infty$ obtained from the pre-symplectic Hamiltonian system above via the pre-symplectic constraint algorithm is symplectic or pre-symplectic.
    \item The (pre-)symplectic structure $\pssos$ can be used to equip the solution space with a Poisson structure.
    In particular, in \cite{Ciaglia-DC-Ibort-Marmo-Schiav-Zamp2021-Cov_brackets_toappear, Ciaglia-DC-Ibort-Mar-Schiav-Zamp2022-Non_abelian} we saw that the structure turns out to be pre-symplectic when the theory exhibits gauge symmetries and that in this case the coisotropic embedding theorem can be used as a tool to define a Poisson bracket from such a pre-sympplectic structure.
\end{itemize}

\subsection{Lagrange multipliers theorem}
\label{Subsec:Lagrange multipliers theorem}

Some theories, such as the main object of the present paper, i.e. General Relativity in the Palatini's formulation, can be described by searching for the extrema of $\ac$ in the Schwinger-Weiss variational principle not in the whole $\fpe$ but rather within a subset of it, say $\Xi$.
Here we discuss whether and how one can use Lagrange's multipliers theorem to search for extrema restricted to $\Xi$.

To deal with this situation in the example we will consider along the manuscript it is sufficient to recall the following version of the Lagrange multipliers theorem on Banach spaces. 
\begin{theorem}[\textsc{Lagrange multipliers theorem}] \label{Prop:lagrange multipliers}
Let $\mathcal{M}$ be a Banach space and let $\mathscr{F}$ be a real-valued differentiable function on $\mathcal{M}$.
Let $\mathcal{N}$ be a Banach space and $\Phi$ a smooth injective map from $\mathcal{N}$ to $\mathcal{C} \,=\, \Phi(\mathcal{N})$ with non-degenerate tangent map. 
Let $\mathcal{M}^\star$ denote the dual of $\mathcal{M}$ and let us define the real-valued function $\mathscr{F}^\mathrm{ext}$ on $\mathcal{M} \times \mathcal{M}^\star \times \mathcal{N}$:
\be
\mathscr{F}^{\mathrm{ext}}_{(m,\Lambda, n)} \,=\, \mathscr{F}_m + \langle \Lambda,\, m - \Phi(n) \rangle
\ee
where $\Lambda$ represents a point in $\mathcal{M}^\star$ and $\langle \,\cdot \,,\, \cdot \,\rangle$ represents the pairing between $\mathcal{M}$ and its dual.

Then, $m$ is a critical point for $\mathscr{F}\bigr|_{\mathcal{C}}$ iff $(m, \Lambda, n)$ is a critical point of $\mathscr{F}^\mathrm{ext}$.
\begin{proof}
First, let us prove that if $m \in \mathcal{C}$ is a critical point for $\mathscr{F}$ then, there exist $\Lambda \in \mathcal{M}^\star$ and $n \in \mathcal{N}$ such that $(m,\, \Lambda,\, n)$ is a critical point for $\mathscr{F}^\mathrm{ext}$.
The $n$ is determined by the fact that, since $\Phi$ is injective and $m$ is in the image of $\Phi$, then there exists a unique $n \in \mathcal{N}$ such that $m = \Phi(n)$.
What is more, since the tangent map of $\Phi$ is non-degenerate, for any $X_n \in \mathbf{T}_n \mathcal{N}$ there exists a $X_m \in \mathbf{T}_{\Phi(n)} \mathcal{C}$ such that $\Phi_\star X_n \,=\, X_m$.
Now, since $m \in \mathcal{C}$ is critical for $\mathscr{F}$, then $\dd \mathscr{F}_m\bigr|_\mathcal{C} \,=\,0$.
On the other hand denoting by $X_m$, $X_\Lambda$ and $X_n$ the components of a tangent vector $X_{(m, \Lambda, n)} \in \mathbf{T}_{(m, \Lambda, n)}\mathcal{M} \times \mathcal{M}^\star \times \mathcal{N}$, $\dd \mathscr{F}^\mathrm{ext}$ is computed to be:
\be \label{Eq:variation extended action}
\dd \mathscr{F}^\mathrm{ext}_{(m,\Lambda, n)}(X_{(m, \Lambda, n)}) \,=\, \dd \mathscr{F}_m(X_m) + \langle X_\Lambda,\, m - \Phi(n) \rangle + \langle \Lambda,\, X_m - \Phi_\star X_n \rangle \,. 
\ee
Now, the three terms on the right hand side of the latter equation all vanish because $\dd \mathscr{F}_m \bigr|_{\mathcal{C}} \,=\, 0$, $m \,=\, \Phi(n)$ and $X_m \,=\, \Phi_\star X_n$ (i.e. $X_m$ and $X_n$ are $\Phi$-related).

Now, let us prove the converse, that is, that if $(m, \Lambda, n)$ is a critical point for $\mathscr{F}^\mathrm{ext}$ then $m$ is a critical point for $\mathscr{F}$.
Since $(m, \Lambda, n)$ is a critical point for $\mathscr{F}^\mathrm{ext}$ then $\dd \mathscr{F}^\mathrm{ext}_{(m, \Lambda, n)}(X_{(m, \Lambda, n)}) = 0$ for all $X_{(m, \Lambda, n)}$.
In particular, if we consider a $X_{(m, \Lambda, n)}$ only having component $X_\Lambda$, then Eq. \eqref{Eq:variation extended action} gives $m = \Phi(n)$.
This tells us that critical points of $\mathscr{F}^\mathrm{ext}$ are such that $m$ is the image via $\Phi$ of some $n \in \mathcal{N}$, i.e., $m \in \mathcal{C}$.
Now, since $\Phi$ is injective and its tangent map is non-degenerate then for any $X_n \in \mathbf{T}_n \mathcal{N}$ there exists $X_m \in \mathbf{T}_m \mathcal{C}$ such that $X_m \,=\, \Phi_\star X_n$.
Therefore, tangent vectors to critical points of $\mathscr{F}^\mathrm{ext}$ are such that the components $X_m$ and $X_n$ are related by the equality $X_m \,=\, \Phi_\star X_n$.
Consequently, by looking at Eq. \eqref{Eq:variation extended action} we get that if $(m, \Lambda, n)$ is a critical point for $\mathscr{F}^\mathrm{ext}$, then $\dd \mathscr{F}^\mathrm{ext}_{(m,\Lambda, n)} \,=\, \dd \mathscr{F}_m \,=\, 0$.
\end{proof}
\end{theorem}

By looking at our situation, if the subset of fields $\Xi$ where we want to search for extrema of $\ac$ is the image into $\fpe$ of some Banach space $\mathcal{N}$ via a map $\Phi$ satisfying the hypothesis of the latter proposition, then we can search for the extrema of $\ac$ restricted to $\Xi$ by searching for the extrema of the functional $\ac^{\mathrm{ext}}$ defined on $\fpe \times \fpe^\star \times \mathcal{N}$:
\be
\ac^{\mathrm{ext}}_{(\chi, \Lambda, n)} \,=\, \ac_\chi + \langle \Lambda,\, \chi - \Phi(n) \rangle \,.
\ee
Explicitly, in the system of local coordinates chosen in \cite{Ciaglia-DC-Ibort-Marmo-Schiav-Zamp2021-Cov_brackets_toappear, Ciaglia-DC-Ibort-Mar-Schiav-Zamp2022-Non_abelian}, the term $\langle \Lambda,\, \chi - \Phi(n) \rangle$ reads:
\be 
\langle \Lambda,\, \chi - \Phi(n) \rangle \,=\, \int_\m \left[\, \Lambda^\phi_a \left( \phi^a - u^a \circ \Phi(n) \right) + {\Lambda^P}^a_\mu \left( P^\mu_a - \rho^\mu_a \circ \Phi(n) \right) \,\right] vol_\m \,,
\ee
where $(\Lambda^\phi_a ,\, {\Lambda^P}^\mu_a)$ is a system of local coordinates on $\fpe^\star$.
Consequently, $\ac^{\mathrm{ext}}$ explicitly reads:
\be \label{Eq:extended action functional explicit}
\ac^{\mathrm{ext}}_{(\chi, \Lambda, n)} \,=\, \int_\m \left[\, P^\mu_a \de_\mu \phi^a - H(\chi) + \Lambda^\phi_a \left( \phi^a - u^a \circ \Phi(n) \right) + {\Lambda^P}^a_\mu \left( P^\mu_a - \rho^\mu_a \circ \Phi(n) \right) \,\right] vol_\m \,.
\ee


\subsection{The canonical structure on the space of solutions}
\label{Subsec:The canonical structure on the space of solutions}

By looking at Eq. \eqref{Eq:variation extended action} we see that the additional term appearing in the variation of $\ac^{\mathrm{ext}}$ in this "constrained situation" reads:
\be
\langle X_\Lambda,\, m - \Phi(n) \rangle + \langle \Lambda,\, X_m - \Phi_\star X_n \rangle \,,
\ee
which is not a boundary term and, thus, does not contribute to the one-form $\Pi^\star_\Sigma \alpha^\Sigma$ defined in \cite{Ciaglia-DC-Ibort-Marmo-Schiav-Zamp2021-Cov_brackets_toappear}.
This means that the canonical structure $\pssos$ associated to $\ac^{\mathrm{ext}}$ has the same expression as the one associated with $\ac$ even if defined on the enlarged space $\fpe \times \fpe^\star \times \mathcal{N}$.

On the other hand, as it is clear from Eq. \eqref{Eq:extended action functional explicit}, the additional term $\langle \Lambda,\, \chi - \Phi(n) \rangle$ in $\ac^{\mathrm{ext}}$ has the net result of subtracting to the Hamiltonian $H$, the function $\Lambda^\phi_a \left( \phi^a - u^a \circ \Phi(n) \right) + {\Lambda^P}^a_\mu \left( P^\mu_a - \rho^\mu_a \circ \Phi(n) \right)$.
Consequently, the analogue of the Hamiltonian functional defined in \cite{Ciaglia-DC-Ibort-Marmo-Schiav-Zamp2021-Cov_brackets_toappear} obtained from $\ac^{\mathrm{ext}}$ turns out to be modified in the following way:
\be \label{Eq:hamiltonian functional multipliers}
\begin{split}
\mathcal{H}^\mathrm{ext}(\chi_\Sigma,\, \Lambda_\Sigma, n) \,&=\, \int_\Sigma \biggl\{\, \beta^k_a \de_k \varphi^a - H(\chi_\Sigma) + \\ 
&+ \left[ \lambda^\varphi_a \left( \varphi^a - u^a \circ \Phi(n)\bigr|_\Sigma \right) + {\lambda^P}^a_0 \left( p_a - \rho^0_a \circ \Phi(n)\bigr|_\Sigma \right) + {\lambda^P}^a_k \left( \beta^k_a - \rho^k_a \circ \Phi(n)\bigr|_\Sigma \right) \right] \,\biggr\} vol_\Sigma \,,
\end{split}
\ee
where $\lambda^\varphi_a \,:=\, \Lambda^\phi_a\bigr|_\Sigma$ and ${\lambda^P}^\mu_a \,=:\, {\Lambda^P}^\mu_a\bigr|_\Sigma$.
Therefore, in this case the pre-symplectic Hamiltonian system associated to the modified action, $\ac^\mathrm{ext}$, is $\left(\, \fpe^\Sigma \times {\fpe^\Sigma}^\star \times \mathcal{N},\, {\os}^\mathrm{ext},\, \mathcal{H}^\mathrm{ext}  \,\right)$ where ${\os}^\mathrm{ext} \,=\, \tau^\star \os$, $\tau$ being the projection $\tau \; :\; \fpe^\Sigma \times {\fpe^\Sigma}^\star \times \mathcal{N} \to \fpe^\Sigma$.
With this in mind, \cite[Prop. $4.2$]{Ciaglia-DC-Ibort-Marmo-Schiav-Zamp2021-Cov_brackets_toappear}, in this case is generalized as follows.
\begin{proposition}
    Extrema of the functional ${\tilde{\ac}^\mathrm{ext}}^\epsilon$ are the solutions of the pre-symplectic system $\left(\, \fpe^\Sigma \times {\fpe^\Sigma}^\star \times \mathcal{N},\, {\os}^\mathrm{ext},\, \mathcal{H}^\mathrm{ext}  \,\right)$.
\end{proposition} 


\section{Palatini's formulation of General Relativity}
\label{Sec:Palatini's formulation of General Relativity}

In this section we use the theory developed in Sec. \ref{Subsec:Lagrange multipliers theorem} and \ref{Subsec:The canonical structure on the space of solutions} to show how to equip the space of solutions of Einstein's equations with a Poisson structure.
In particular, in Sec. \ref{Subsec:Variational formulation of General Relativity: from Einstein-Hilbert to Palatini} and \ref{Subsec:The topological limit and the Palatini's constraint} we show how General Relativity can be formulated as a particular case of Yang-Mills theory in a suitable limit that we will call \textit{topological limit} and by imposing a suitable constraint that we call \textit{Palatini's constraint}.
In Sec. \ref{Subsec:Schwinger-Weiss variational principle and the formalism near a slice} we will explicit construct the pre-symplectic structure on the space of solutions of (torsionless) Einstein's equations in vacuum via the pre-symplectic constraint algorithm.
Finally, in Sec. \ref{Subsec:Covariant brackets and DeWitt's formula} we will argue how, provided with the existence of a flat connection on a suitable bundle emerging in Sec. \ref{Subsec:Schwinger-Weiss variational principle and the formalism near a slice} (that we conjecture), the results of Sec. \ref{Subsec:Schwinger-Weiss variational principle and the formalism near a slice} can be used to construct a Poisson bracket as it was done in \cite{Ciaglia-DC-Ibort-Marmo-Schiav-Zamp2021-Cov_brackets_toappear, Ciaglia-DC-Ibort-Mar-Schiav-Zamp2022-Non_abelian}.

\subsection{Variational formulation of General Relativity: from Einstein-Hilbert to Palatini}
\label{Subsec:Variational formulation of General Relativity: from Einstein-Hilbert to Palatini}

General Relativity can be described as a field theory over a space-time $(\m,\, g)$ where the metric $g$ is actually the configuration field of the theory.
Historically, the first variational formulation of the theory is due to \textit{A. Einstein} and \textit{D. Hilbert}.
It can be proved \cite[Chap. 3]{Baez1994-Gauge_theories}, that Einstein equations can be obtained as Euler-Lagrange equations for the following action functional:
\be
{\ac_{\textsc{E-H}}}_g \,=\, \int_\m \mathcal{R} \, \epsilon \, vol_\m \,,
\ee
which is called the \textit{Einstein-Hilbert action}, where $\mathcal{R}$ is the scalar curvature of $g$, $\epsilon \,=\, \sqrt{-\mathrm{det}g}$ and $vol_{\m} \,=\, \dd x^0 \wedge ... \wedge \dd x^d$.

We briefly recall how $\mathcal{R}$ is defined because it will be useful when we will pass to the tetradic formalism.
Being $\mathbf{T}\m$ a vector bundle, a linear connection can always be defined on it \cite[Appendix $11.4$]{Giac-Mang-Sard2010-Geometric_Classical_Quantum_Mechanics}.
It is a decomposition of the tangent space to $\mathbf{T}\m$ at each point into the vertical subspace and a horizontal one.
Such a decomposition can be encoded into a vertical valued one-form on $\mathbf{T}\m$ of the type:
\be
A \,=\, \left(\, \dd v^\mu - A^{\ \mu}_{\nu \ \rho} v^\rho \dd x^\nu \,\right) \otimes \frac{\de}{\de v^\mu} \,,
\ee
where $(x^\mu,\, v^\mu)_{\mu=0,...,d}$ is a local coordinate system on $\mathbf{T}\m$ defined on an open neighborhood $U_{\mathbf{T}\m} \subset \mathbf{T}\m$ and $A^{\ \mu}_{\nu \ \rho}$ are the so-called \textit{connection coefficients}.
The curvature of the connection $\Gamma$ is defined to be the Frolicher-Nijenhuis bracket between $A$ and itself:
\be \label{Eq:curvature}
R \,:=\, [A,\, A]_{F-N} \,=\, R_{\lambda \mu \ \rho}^{\ \ \ \nu} v^\rho \dd x^\lambda \wedge \dd x^\mu \otimes \frac{\de}{\de v^\nu} \,,
\ee
where:
\be
R_{\lambda \mu \ \rho}^{\ \ \ \nu} \,=\, \frac{1}{2} \left(\, \de_\lambda A_{\mu \ \rho}^{\ \nu} - \de_\mu A_{\lambda \ \rho}^{\ \nu} + A_{\lambda \ \rho}^{\ \sigma} A_{\mu \ \sigma}^{\ \nu} - A_{\mu \ \rho}^{\ \sigma} A_{\lambda \ \sigma}^{\ \nu} \,\right) \,.
\ee
The Ricci tensor, $\mathscr{R}$, of the connection $A$ is the $(0,2)$-tensor on $\m$ whose coefficients are $\mathscr{R}_{\mu \nu} \,=\, R_{\mu \rho \ \nu}^{\ \ \ \rho}$ and the scalar curvature $\mathcal{R}$ is:
\be
\mathcal{R} \,=\, g^{\mu \nu} \mathscr{R}_{\mu \nu} \,.
\ee

In the Einstein-Hilbert formulation, General Relativity is a second order theory, in the sense that the action functional depends on the fields and their derivatives up to second order, since $\mathcal{R}$ actually depends on $g$ and its first and second order derivatives.
Even if a quite solid geometrical theory to work with higher order field theories exists (see \cite{Krupka2015-Variational_Geometry} and references therein), it is much easier to work with first order theories when possible.
Indeed, General Relativity can be put into a first order theory by considering the metric $g$ and the connection $A$ as independent objects.
This is the so-called \textit{Palatini's variational formulation} of General Relativity.
Palatini's action reads:
\be \label{Eq:palatini action}
{\ac_{\textsc{P}}}_{(g,A)} \,=\, \int_\m g^{\mu \nu} \mathscr{R}_{\mu \nu} \, \epsilon \, vol_\m \,,
\ee
where $\mathscr{R}$ is the Ricci curvature of the connection $A$.

There are several good reasons for not taking $g$ as the fundamental field of the theory, but rather the so-called \textit{tetrad fields}.
A tetrad at a point $m \in \m$ is defined to be a map from a basis of the tangent space of $\m$ at $m$ to elements of the tangent space of a flat $n$-dimensional\footnote{Actually, even if the definition does not depend on the dimension, the name \textit{tetrad} is devoted to the $4$-dimensional case, which is the case we will address.} manifold (which is isomorphic with $\mathbb{R}^n$) equipped with a Lorentzian metric:
\be 
e(x) \;\;: \;\; \mathbf{T}_m \m \to \mathbb{R}^n \;\; : \;\; \frac{\de}{\de x^\mu}  \mapsto  e(x)\left(\frac{\de}{\de x^\mu}\right) \,=\, e_\mu^I(x) \xi_I \,, \;\;\; I \,=\, 1,...,n \,,
\ee
where $\xi_I$ is a basis of the vector space $\mathbb{R}^n$.
Therefore, $e(x)$ can be thought of as a $1-1$ tensor on $\mathbf{T}_m \m \otimes \mathbb{R}^n$ of the type:
\be
e(x) \,=\, e_\mu^I(x) \dd x^\mu \otimes \xi_I \,,
\ee
which satisfies:
\be \label{Eq:tetrad condition}
g_{\mu \nu} e^\mu_I e^\nu_J \,=\, \eta_{IJ} \,,
\ee
$\eta$ being the Minkowski metric which $\mathbb{R}^n$ is equipped with.
The existence of such a map is ensured by the existence around any point $m$ of $\m$ of normal (or Gauss) coordinates in which the metric $g$ is the Minkowski metric and its first derivatives in $m$ vanish (see \cite{Spivak1999-Vol2}).
Now, a tetrad field is defined to be a local section $e$ of the bundle $\mathbf{T}\m \otimes \mathbb{R}^n \to \m$ satisfying, at each point, a condition of the type \eqref{Eq:tetrad condition}.

The duals of the tetrad fields are defined by considering the following map:
\be
e^\star(x) \;\;: \;\; \mathbf{T}^\star_m \m \to \mathbb{R}^n \;\; : \;\; \dd x^\mu  \mapsto  e^\star(x)\left(\dd x^\mu \right) \,=\, e^\mu_I(x) \xi^I \,, \;\;\; I \,=\, 1,...,n \,,
\ee
where $\left\{\, \xi^I \,\right\}_{I = 1,...,n}$ is the dual basis of $\left\{\, \xi_I \,\right\}_{I = 1,...,n}$.
$e^\star(x)$ can be thought of as a $1-1$ tensor on $\mathbf{T}^\star_m \m \otimes \mathbb{R}^n$ of the type:
\be
e^\star(x) \,=\, e^\mu_I(x) \frac{\de}{\de x^\mu} \otimes \xi^I \,,
\ee
which satisfies:
\be \label{Eq:dual tetrad condition}
g^{\mu \nu} e_\mu^I e_\nu^J \,=\, \eta^{IJ} \,.
\ee
Therefore, a dual tetrad field is defined to be a local section of the bundle $\mathbf{T}^\star \m \otimes \mathbb{R}^n \to \m$ satisfying, at each point, the condition \eqref{Eq:dual tetrad condition}.

\begin{remark}
The use of the $e$'s instead of $g$ as fundamental field of the theory is widespread nowadays.
Indeed, within the most common approaches to Quantum Gravity the use of tetrads is necessary to describe fermions.
Moreover, tetrads describe very well the idea of the gravitational field as a deviation of the space-time from being flat and, therefore, many authors prefer to use $e$ instead of $g$ also from the point of view of the sake of Physical conceptual clearness (see \cite{Rovelli2004-Quantum_Gravity}).
\end{remark}

Now, the action $\ac_{\textsc{P}}$ can be expressed in terms of the connection $A$ and the tetrad fields in the following way.
The curvature \eqref{Eq:curvature} can be expressed in the basis of the $\xi_I$'s by taking the pull-back of $R$ via the inverse of the map $e^\star$:
\be
\left({e^\star}^{-1}\right)^\star R \,=\, R_{\lambda \mu \ \rho}^{\ \ \ \nu} e^\lambda_I e^\mu_J v^\rho \xi^I \wedge \xi^J \otimes \frac{\de}{\de v^\nu} \,=:\, R_{IJ \ \rho}^{\ \ \ \nu}v^\rho \xi^I \wedge \xi^J \otimes \frac{\de}{\de v^\nu} \,.
\ee
Consequently, the Ricci tensor can be expressed in terms of the tetrad fields as follows:
\be
\mathscr{R}_{\mu \nu} \,=\, R_{\mu \sigma \ \nu}^{\ \ \ \sigma} \,=\, R_{IJ \ \nu}^{\ \ \ \sigma} e^I_\mu e^J_\sigma \,,
\ee
and the scalar curvature reads:
\be
\mathcal{R} \,=\, g^{\mu \nu} \mathscr{R}_{\mu \nu} \,=\, \eta^{IJ}e^\mu_I e^\nu_J R_{KL \ \nu}^{\ \ \ \sigma} e^K_\mu e^L_\sigma \,=\, - e^\nu_K e^\sigma_L R^{KL}_{\ \ \ \nu \sigma} \,.
\ee
Therefore, in terms of tetrad fields, the action \ref{Eq:palatini action} reads:
\be \label{Eq:tetradic palatini}
{\ac_{\textsc{P}}}_{(e, A)} \,=\, - \int_\m \epsilon e^\mu_I e^\nu_J R^{IJ}_{\ \ \ \mu \nu} vol_\m \,. 
\ee
which is the so-called \textit{tetradic Palatini's action}.

\begin{remark} \label{Rem:curvature o(1,3)}
It is worth noting that $R^{IJ}_{\ \ \ \mu \nu}$ are the coefficient of a $2$-form on $\m$ with values in $(\mathbb{R}^n \wedge \mathbb{R}^n,\, \eta)$.
When $n=4$, $(\mathbb{R}^n \wedge \mathbb{R}^4,\, \eta)$ is isomorphic to the Lie algebra of the orthogonal group $O(1,3)$.
Therefore, the indices $IJ$ can be considered as a collective index $a \,=\, IJ \,=\, 1,...,\mathrm{dim}\mathfrak{o}(1,3)$ and $R^{IJ}_{\ \ \ \mu \nu}$ can be seen as the coefficients of the curvature of a connection one-form on $\m$ with values in $\mathfrak{o}(1,3)$. 
\end{remark}

The main goal of the following section is to develop the multisymplectic formulation of the tetradic Palatini's action.
In particular, we will see that \eqref{Eq:tetradic palatini} can be regarded as a Yang-Mills action of the type introduced in \cite{Ciaglia-DC-Ibort-Mar-Schiav-Zamp2022-Non_abelian} in a suitable limit, the so-called \textit{topological limit}, and constrained to a suitable subset of fields.
For this reason we will necessitate the theory developed in Sec. \ref{Subsec:Lagrange multipliers theorem} to deal with such a theory.

\subsection{The topological limit and the Palatini's constraint}
\label{Subsec:The topological limit and the Palatini's constraint}

First let us consider the Yang-Mills action \cite{Ciaglia-DC-Ibort-Mar-Schiav-Zamp2022-Non_abelian} with a dimensional constant multiplying the quadratic term in the momenta fields:
\be
{\ac_{\textsc{Y-M}}}_\chi \,=\, -\int_\m \left[\, P^{\mu \nu}_a F^a_{\mu \nu} + \frac{1}{4 G} P^{\mu \nu}_a P_{\mu \nu}^a \,\right] vol_\m \,. 
\ee
Its \textit{topological limit}\footnote{The reason for this name is that the theory it will give rise is a topological one, in the sense that the dynamics of the theory will lie entirely into the kernel of a pre-symplectic structure, as it will be clear in the next section.} is defined to be:
\be \label{Eq:topological limit yang-mills}
{\ac_{\textsc{Y-M}}^0}_\chi \,=\, \mathrm{lim}_{G \to \infty} {\ac_{\textsc{Y-M}}}_\chi \,=\, - \int_\m P^{\mu \nu}_a F_{\mu \nu}^a \, vol_\m \,.
\ee
Note that such a topological limit is obtained by considering the limit for $G \to \infty$ of the Yang-Mills Hamiltonian with a dimensional constant $G$:
\be
H \,=\, \frac{1}{4 G} \rho^{\mu \nu}_a \rho^{a}_{\mu \nu} + \frac{1}{2} \epsilon^a_{bc} \rho^{\mu \nu}_a \alpha^b_\mu \alpha^c_\nu \,.
\ee

Having in mind Remark \ref{Rem:curvature o(1,3)}, the action \eqref{Eq:tetradic palatini} can be seen as the topological limit of the action of a Yang-Mills theory with structure group $O(1,3)$ with the identification of the momenta of the theory $P^{\mu \nu}_a$ with the expression $\epsilon e^\mu_I e^\nu_J$ appearing in Eq. \eqref{Eq:palatini action}.
Let us make this last claim more precise.

Having in mind Remark \ref{Rem:curvature o(1,3)}, the internal indices of a $O(1,3)$ Yang-Mills configuration fields can be written as $a \,=\, IJ$ where $I, J$ runs from $0$ to $3$ and $IJ$ must be considered as a collective index running on $\{0,...,3\} \wedge \{0,...,3\}$.
Therefore, here we denote Yang-Mills configuration fields as:
\be
A \,=\, A(x)_\mu^{IJ} \dd x^\mu \otimes \xi_I \wedge \xi_J \,,
\ee
where $\{\,\xi_I \,\}_{I=0,...,3}$ represents a basis of $\mathbb{R}^4$, $\{\, \xi_I \wedge \xi_J \,\}_{I,J=0,...,3}$ represents a basis of $\mathfrak{o}(1,3)$.
On the other hand, momenta fields will be denoted as:
\be
P \,=\, P^{\mu \nu}_{IJ}(x) \frac{\de}{\de x^\mu} \wedge \frac{\de}{\de x^\nu} \otimes \xi^I \wedge \xi^J \,,
\ee
where $\{\, \xi^I \,\}_{I=0,...,3}$ is a basis of $\mathbb{R}^4$, $\{\, \xi^I \wedge \xi^J \,\}_{I,J=0,...,3}$ represents a basis of $\mathfrak{o}(1,3)^\star$.
As usual, dynamical fields of the theory are denoted by $\chi \,=\, (A,\, P) \in \fpe$.
How to give it the structure of a Banach space is extensively discussed in \cite{Ciaglia-DC-Ibort-Mar-Schiav-Zamp2022-Non_abelian}.
On the other hand, we saw that tetrad fields are defined as sections of the bundle $\mathbf{T}\m \otimes \mathbb{R}^4 \to \m$ which reads:
\be
e \,=\, e^\mu_I(x) \frac{\de}{\de x^\mu} \otimes \xi^I \,.
\ee
Let us denote by $\mathscr{E}$ the space of tetrad fields.
Differently to what we did for the space $\fpe$, to which we carefully gave the structure of a Banach manifold in \cite{Ciaglia-DC-Ibort-Mar-Schiav-Zamp2022-Non_abelian}, here we proceed in a formal way assuming that $\mathscr{E}$ is a Banach manifold as well and we postpone a careful functional-analytic analysis to future works.
Then, the following map can be defined:
\be
\mathscr{P} \;\; : \;\; \mathcal{F}_{\mathbb{E}} \times \mathscr{E} \to \fpe \;\; :\;\; (A,\,e) \mapsto \left(\, A,\, P \,=\, \epsilon e^\mu_I e^\nu_J \frac{\de}{\de x^\mu} \wedge \frac{\de}{\de x^\nu} \otimes \xi^I \wedge \xi^J \,\right) \,.
\ee
We will refer to it as \textit{Palatini map} or \textit{Palatini constraint}.
Then, the tetradic Palatini's action is nothing but the pull-back via $\mathscr{P}$ of the topological limit of the Yang-Mills action:
\be
{\ac_{\textsc{P}}} \,=\, \mathscr{P}^\star {\ac_{\textsc{Y-M}}^0} \,,
\ee
whose extrema are the extrema of $\ac^0_{\textsc{Y-M}}$ constrained along the image of $\mathscr{P}$.
Therefore, being $\mathscr{P}$ a map satisfying the hypothesis of Prop. \ref{Prop:lagrange multipliers}, we are in the situation depicted in Sec. \ref{Subsec:Lagrange multipliers theorem} where $\fpe$ is the space of dynamical fields of a $O(1,3)$ Yang-Mills theory and $\Xi$ is the subset of $\fpe$ being the image of $\mathscr{E}$ via the map $\Phi \,=\, \mathscr{P}$.
Therefore, in the next section, we will find extrema of $\ac_{\textsc{P}}$ by using the theory developed in Sect. \ref{Subsec:Lagrange multipliers theorem} and the formalism near a slice $\Sigma$ in $\m$ developed in Sect. \ref{Subsec:The canonical structure on the space of solutions}.

\subsection{Schwinger-Weiss variational principle and the formalism near a slice}
\label{Subsec:Schwinger-Weiss variational principle and the formalism near a slice}

Following the theory developed in Sec. \ref{Subsec:Lagrange multipliers theorem}, extrema of the tetradic Palatini's action \eqref{Eq:tetradic palatini}, that coincide with the extrema of the topological Yang-Mills action \eqref{Eq:topological limit yang-mills} restricted to the image of the Palatini map $\mathscr{P}$, are in one-to-one correspondence with extrema of the action ${\ac^0_{\textsc{Y-M}}}^{\mathrm{ext}}$ defined on $\fpe \times \fpe^\star \times \mathscr{E}$:
\be
{\ac^0_{\textsc{Y-M}}}^{\mathrm{ext}}_{(\chi,\, \Lambda,\, e)} \,=\, \int_\m \left[\, -P^{\mu \nu}_{IJ} R^{IJ}_{\ \ \ \mu \nu} + {\Lambda^P}^{IJ}_{\mu \nu} \left(\, P^{\mu \nu}_{IJ} - \epsilon e_I^\mu e^J_\nu \,\right) \,\right] vol_\m \,,
\ee
where $({\Lambda^A}_{IJ}^\mu,\, {\Lambda^P}^{IJ}_{\mu \nu})$ is a system of local coordinates on $\fpe^\star$.
Since the map $\mathscr{P}$ does not involve the $A$ fields, only the $P$ components of $\Lambda$ always will appear in the computations.
Therefore, without any danger of confusion, we will denote ${\Lambda^P}^{IJ}_{\mu \nu}$ simply as $\Lambda^{IJ}_{\mu \nu}$.

Now, following what we saw in Sec. \ref{Subsec:The canonical structure on the space of solutions}, for any slice $\Sigma$ of $\m$, the extrema of ${\ac^0_{\textsc{Y-M}}}^{\mathrm{ext}}$, near $\Sigma$ are the solutions of the pre-symplectic Hamiltonian system $\left( \fpe^\Sigma \times {\fpe^\Sigma}^\star \times \mathscr{E}, {\Omega^\Sigma}^{\mathrm{ext}},\, \mathcal{H}^{\mathrm{ext}}  \right)$ where ${\os}^{\mathrm{ext}} \,=\, \tau^\star \Omega^\Sigma$ with $\tau$ being the projection $\fpe^\Sigma \times {\fpe^\Sigma}^\star \times \mathscr{E} \to \fpe^\Sigma$ and $\os$ being the canonical pre-symplectic structure on the Yang-Mills fields restricted to $\Sigma$ appearing in \cite{Ciaglia-DC-Ibort-Mar-Schiav-Zamp2022-Non_abelian} and where (having in mind \eqref{Eq:hamiltonian functional multipliers}) $H^{\mathrm{ext}}$ reads:
\be
\mathcal{H}^{\mathrm{ext}}(\overline{\chi}_\Sigma) \,=\, \int_\Sigma \left[\, p^k_{IJ} \left(\, \nabla_k a_0^{IJ} - 2 \lambda^{IJ}_{k0} \, \right) + \beta^{IJ}_{jk} \left(\, F_{jk}^{IJ} - \lambda^{IJ}_{jk} \,\right) + 2 \lambda^{IJ}_{k0} \, \epsilon e^k_I e^0_J  - \lambda^{IJ}_{jk} \, \epsilon e^j_I e^k_J \,\right] vol_\Sigma \,,
\ee
where $\overline{\chi}_\Sigma \,=\, (\chi_\Sigma,\, \Lambda_\Sigma,\, e)$ and $\lambda^{IJ}_{\mu \nu} \,=\, \Lambda^{IJ}_{\mu \nu}\bigr|_{\Sigma}$.

Solutions of such a pre-symplectic Hamiltonian system can be obtained by applying the pre-symplectic constraint algorithm already used in the first two parts of this series of papers.
Denoting by $\mathcal{M} \,=\, \fpe^\Sigma \times {\fpe^\Sigma}^\star \times \mathscr{E}$, the first step of the algorithm gives the first constraint manifold:
\be \label{Eq:pca 1 palatini}
\mathfrak{i}_1\left(\mathcal{M}_1\right) \,=\, \left\{\, \overline{\chi}_\Sigma \in \fpe^\Sigma \times {\fpe^\Sigma}^\star \times \mathscr{E} \;\; :\;\; i_{\mathbb{X}_{\overline{\chi}_\Sigma}} \dd \mathcal{H}^{\mathrm{ext}}_{\overline{\chi}_\Sigma} \,=\, 0 \;\; \forall \,\, \mathbb{X}_{\overline{\chi}_\Sigma} \in \mathbf{T}_{\overline{\chi}_\Sigma}\mathcal{M}^\perp \,\right\} \,,
\ee
where:
\be
\mathbf{T}_{\overline{\chi}}\mathcal{M}^\perp \,=\, \left\{\, \mathbb{X}_{\overline{\chi}_\Sigma} \in \mathbf{T}_{\overline{\chi}_\Sigma}\mathcal{M} \;\; :\;\; i_{\mathbb{X}_{\overline{\chi}_\Sigma}} {\os}^{\mathrm{ext}} \,=\, 0 \,\right\} \,,
\ee
is just the kernel of ${\os}^{\mathrm{ext}}$ at $\overline{\chi}_\Sigma$.
By looking at the expression of ${\os}^{\mathrm{ext}}$ (see \cite{Ciaglia-DC-Ibort-Mar-Schiav-Zamp2022-Non_abelian}), the kernel is seen to be generated by tangent vectors having only component $a_0^{IJ}$, $\beta^{IJ}_{jk}$, $e^0_I$, $e^j_I$, $\lambda_{jk}^{IJ}$, $\lambda_{k0}^{IJ}$ respectively.
We will denote by:
\be
\left\{\, \frac{\delta}{\delta a^{IJ}_0},\, \frac{\delta}{\delta \beta^{jk}_{IJ}},\, \frac{\delta}{\delta e^0_I},\, \frac{\delta}{\delta e^k_I},\,  \frac{\delta}{\delta \lambda^{jk}_{IJ}},\, \frac{\delta}{\delta \lambda^{0k}_{IJ}} \,\right\}
\ee
a basis of $\mathbf{T}_{\overline{\chi}_\Sigma} \mathcal{M}^\perp$.
Then, the condition in \eqref{Eq:pca 1 palatini} gives the following constraints:
\be \label{Eq:constraints palatini}
\begin{split}
\nabla_k p^k_{IJ} \,=\, 0 \,, \qquad F^{IJ}_{jk} \,&=\, \lambda^{IJ}_{jk},\, \qquad p^k_{IJ} \,=\, \epsilon e^0_I e^k_J \,, \\
\beta_{jk}^{IJ} \,=\, \epsilon e^j_I e^k_J \,, \qquad e^k_I \lambda^{IJ}_{k0} \,&=\,0 \,, \qquad e^0_J \lambda^{IJ}_{k0} \,=\, e^j_J \lambda^{IJ}_{kj} \,.
\end{split}
\ee
These constraints allows to eliminate the fields $a_0^{IJ}$, $p^k_{IJ}$, $\beta^{IJ}_{jk}$ and $\lambda^{IJ}_{jk}$ and, therefore, the manifold $\mathcal{M}_1$ is:
\be
\mathcal{M}_1 \,=\, \left\{\, \left( a_k^{IJ},\, e^\mu_I ,\, \lambda^{IJ}_{jk} \right) \;\; :\;\; \nabla_k p^k_{IJ} \,=\, 0,\,\,\, e^k_I \lambda^{IJ}_{k0} \,=\, 0 ,\,\,\, e^0_J \lambda^{IJ}_{k0} \,=\, e^j_J \lambda^{IJ}_{kj}   \,\right\} \,,
\ee
and its immersion into $\mathcal{M}$ is given by the remaining conditions:
\be
F^{IJ}_{jk} \,=\, \lambda^{IJ}_{jk},\, \qquad p^k_{IJ} \,=\, \epsilon e^0_I e^k_J \,, \qquad \beta_{jk}^{IJ} \,=\, \epsilon e^j_I e^k_J
\ee
and by fixing any arbitrary $a_0^{IJ}$.
It is a matter of direct computation to show that $\mathfrak{i}_1^\star \mathcal{H}^{\mathrm{ext}} \,=\, 0$.
The second step of the PCA gives:
\be
\mathfrak{i}_2(\mathcal{M}_2) \,=\, \left\{\, \overline{\chi}_\Sigma \in \mathcal{M}_1 \;\; :\;\; i_{{\mathbb{X}_{\overline{\chi}}}_\Sigma} \dd \mathcal{H}^{\mathrm{ext}} \,=\, 0 \,\right\} \,, 
\ee
where:
\be
\mathbf{T}_{\overline{\chi}_{\Sigma}}\mathcal{M}_1^\perp \,=\, \left\{\, \mathbb{X}_{\overline{\chi}_\Sigma} \in \mathbf{T}_{\overline{\chi}_\Sigma} \mathcal{M} \;\; : \;\; \mathfrak{i}_1^\star \left(\, i_{\mathbb{X}_{\overline{\chi}_\Sigma}} {\os}^{\mathrm{ext}} \,\right) \,=\, 0 \,\right\} \,.
\ee
$\mathbf{T}_{\overline{\chi}_\Sigma}\mathcal{M}_1^\perp$ is spanned by the tangent vector $\mathbb{X}$ with components:
\be \label{Eq:transversal tangent vector}
{\mathbb{X}_a}_k^{IJ} \,=\, \nabla_k a_0^{IJ} - 2 \lambda^{IJ}_{k0} \,, \qquad {\mathbb{X}_p}^k_{IJ} \,=\, - \nabla_j (\epsilon e^j_{[I} e^k_{J]}) - [p^k,\, a_0]_{IJ}
\ee
where the square brackets in the indices means a skew-symmetrization.
A direct computation shows that:
\be
i_{\mathbb{X}} \dd \mathcal{H}^{\mathrm{ext}} \,=\, 0 \,,
\ee
and, thus, $\mathcal{M}_2$ coincide with $\mathcal{M}_1$ and with the final manifold of the PCA.
Therefore, we will denote it by $\mathcal{M}_\infty$.
On the final manifold the canonical equation reads:
\be
\left(\, i_{\mathbb{\Gamma}}{\os}^{\mathrm{ext}} - \dd \mathcal{H}^{\mathrm{ext}} \,\right)_{\mathcal{M}_\infty} \,=\, 0 \,,
\ee
which, since, as we said above, $\mathfrak{i}_1^\star \mathcal{H}^{\mathrm{ext}} \,=\, 0$, reduce to:
\be
\left(\, i_{\mathbb{\Gamma}}{\os}^{\mathrm{ext}}  \,\right)_{\mathcal{M}_\infty} \,=\, 0 \,.
\ee
This means that the dynamics lies, at each point, entirely in $\mathbf{T}_{\overline{\chi}_\Sigma}\mathcal{M}_\infty^\perp$.
Therefore, solutions of the original pre-symplectic system are the integral curves (restricted to $\mathcal{M}_\infty$) of a vector field $\mathbb{\Gamma}$ that, at each point of $\mathcal{M}$ lies in $\mathbf{T}_{\overline{\chi}_\Sigma}\mathcal{M}_\infty^\perp$.

Now, $\mathbf{T}_{\overline{\chi}_\Sigma}\mathcal{M}_\infty^\perp$ contains a part of vectors being tangent to $\mathcal{M}_\infty$ and a part of vectors being tangent to $\mathcal{M}$ but not tangent to $\mathcal{M}_\infty$. 
We will denote them respectively by:
\be 
\mathbf{T}_{\overline{\chi}_\Sigma}\mathcal{M}_\infty^\perp \,=\, {\mathbf{T}_{\overline{\chi}_\Sigma}\mathcal{M}_\infty^\perp}^\parallel \bigcup {\mathbf{T}_{\overline{\chi}_\Sigma}\mathcal{M}_\infty^\perp}^\perp \,.
\ee
We already found the "transversal" part, because it is spanned by the tangent vector $\mathbb{X}$ in \eqref{Eq:transversal tangent vector}.
The integral curves of a vector field which at each point of $\mathcal{M}$ coincides with $\mathbb{X}$ are the solutions of:
\be \label{Eq:einstein 0}
\begin{split}
\frac{d a_k^{IJ}}{ds} \,&=\, \nabla_k a_0^{IJ} - 2 \lambda^{IJ}_{k0} \,,\\ 
\frac{d p^k_{IJ}}{ds} \,&=\, - \nabla_j \left(\epsilon e^j_{[I} e^k_{J]} \right) - \left[p^k,\, a_0 \right]_{IJ} \,.
\end{split}
\ee 
Their restriction to $\mathcal{M}_\infty$ are the solutions of the combination of \eqref{Eq:einstein 0} and \eqref{Eq:constraints palatini}, which reads:
\be \label{Eq:torsionless + einstein}
\nabla_\mu \left(\, \epsilon e^\mu_{[I} e^\nu_{J]} \,\right) \,=\, 0 \,, \qquad e^\mu_I F^{IJ}_{\mu \nu} \,=\,0 \,.
\ee
The first of the latter equations is the condition for the metric $g$ associated with $e$ to be torsion-less, whereas the second set of equations are Einstein equations in the vacuum (see \cite{Rovelli2004-Quantum_Gravity}).

Regarding the remaining part of $\mathbf{T}_{\overline{\chi}_\Sigma}\mathcal{M}_\infty^\perp$, say the tangent one ${\mathbf{T}_{\overline{\chi}_\Sigma}\mathcal{M}_\infty^\perp}^\parallel$, we will show how it is made by the generators of gauge transformations of the theory.
${\mathbf{T}_{\overline{\chi}_\Sigma}\mathcal{M}_\infty^\perp}^\parallel$ is made by tangent vectors to $\mathcal{M}$ along $\mathcal{M}_\infty$, satisfying:
\be
\left( i_{\mathbb{X}_{\overline{\chi}_\Sigma}} {\os}^{\mathrm{ext}} \right)_{\mathcal{M}_\infty} \,=\, \mathfrak{i}_\infty^\star i_{\mathbb{X}_{\overline{\chi}_\Sigma}} {\os}^{\mathrm{ext}} \,=\,0 \,,
\ee
and which are actually tangent to $\mathcal{M}_\infty$.
This means that $\mathbb{X}_{\overline{\chi}_\Sigma}$ is $\mathfrak{i}_\infty$-related with a tangent vector to $\mathcal{M}_\infty$ at $\overline{\chi}_\Sigma \in \mathcal{M}_\infty$, say $\mathbb{X}^\infty_{\overline{\chi}_\Sigma}$, satisfying:
\be
i_{\mathbb{X}^\infty_{\overline{\chi}_\Sigma}} {\os}^{\mathrm{ext}}_\infty \,=\, 0 \,,
\ee
where ${\os}^{\mathrm{ext}}_\infty \,=\, \mathfrak{i}_\infty^\star {\os}^{\mathrm{ext}}$.
Therefore, let us focus for a moment in searching for the kernel of ${\os}^{\mathrm{ext}}_\infty$.
It is a matter of direct computation to prove that the kernel of ${\os}^{\mathrm{ext}}_\infty$ is spanned, at each point, by the infinity (parametrized by Lie algebra-valued functions $\psi^{IJ}$) of tangent vectors $\mathbb{X}_\psi^{\textsc{g}}$ with components:
\be
{\left(\mathbb{X}^{\textsc{g}}_\psi\right)_a}_k^{IJ} \,=\, \nabla_k \psi^{IJ}  \,, \qquad \, {\left(\mathbb{X}^{\textsc{g}}_\psi\right)_e}^\mu_I \,=\, \psi^K_I e^\mu_K \,,
\ee
and by the infinity (parametrized by vector fields on $\Sigma$) of tangent vectors $\mathbb{X}^{\textsc{d}}$ with components:
\be
{\left(\mathbb{X}^{\textsc{d}}_\xi\right)_a}_k^{IJ} \,=\, \xi^j \nabla_k a_j^{IJ} \,, \qquad {\left(\mathbb{X}^{\textsc{d}}_\xi\right)_e}^0_I \,=\, - \xi^j \nabla_j e^0_K \,, \qquad {\left(\mathbb{X}^{\textsc{d}}_\xi \right)_e}^k_I \,=\, -\xi^j \nabla_j e^k_I \,,
\ee
where $\xi^j$ are the components of a vector field on $\Sigma$.
The vector fields which, at each point, coincide with the tangent vectors $\mathbb{X}^{\textsc{g}}$ satisfy:
\be
[\mathbb{X}^{\textsc{g}}_\psi,\, \mathbb{X}_\phi^{\textsc{g}}] \,=\, \mathbb{X}^{\textsc{g}}_{[\psi, \phi]}
\ee
and, therefore, they are a representation of the Lie algebra $\mathfrak{o}(1,3)$ on $\mathcal{M}_\infty$ given by the following action:
\be \label{Eq:gauge palatini}
\begin{split}
a_\mu^{IJ} &\mapsto \psi^{I}_K a^{KL}_\mu {\psi^{-1}}^{LJ} + {\psi^{-1}}^{J}_K \de_\mu \psi^{KJ} \,, \\
e^\mu_I &\mapsto \psi^K_I e^\mu_K \,, \\
\lambda^{IJ}_{jk} &\mapsto \lambda^{IJ}_jk
\end{split}
\ee
which agree with the gauge transformation written in \cite[Chap. 2, page $41$]{Rovelli2004-Quantum_Gravity} (this justifies the notation $\mathbb{X}^{\textsc{g}}$).
On the other hand, the vector fields which, at each point, coincide with the tangent vectors $\mathbb{X}^{\textsc{D}}$ satisfy:
\be
[\mathbb{X}^{\textsc{d}}_\xi,\, \mathbb{X}^{\textsc{d}}_\zeta] \,=\, \mathbb{X}_{[\xi,\zeta]} \,,
\ee
provided that $\xi$ and $\zeta$ are divergenceless and, therefore, they are a representation of the group of volume-preserving diffeomorphisms (which justifies the notation $\mathbb{X}^{\textsc{d}}$) of $\Sigma$ on $\mathcal{M}_\infty$ given by the following action:
\be \label{Eq:diffeo palatini}
\begin{split}
a_k^{IJ} &\mapsto \left(\mathscr{T}_\nabla^\xi a \right)_k^{IJ} \,,\\
e^\mu_I &\mapsto -\left(\mathscr{T}_\nabla^\xi e \right)^\mu_I \,,\\
\lambda^{IJ}_{jk} &\mapsto \lambda^{IJ}_{jk} \,,
\end{split}
\ee
where $\mathscr{T}_\nabla^\xi$ is the parallel transport along the flow of $\xi$ associated with the connection $\nabla$.
The tangent vectors $\mathbb{X}^{\textsc{g}}$ and $\mathbb{X}^{\textsc{d}}$ are $\mathfrak{i}_\infty$-related with the  tangent vectors to $\mathcal{M}$ having, respectively, components:
\be
{\left(\mathbb{X}^{\textsc{g}}_\psi\right)_a}_k^{IJ} \,=\, \nabla_k \psi^{IJ}  \,, \qquad \, {\left(\mathbb{X}^{\textsc{g}}_\psi\right)_p}^k_{IJ} \,=\, \epsilon \left(\, U_{[I}^L \psi_L^K e^0_K e^k_{J]} - W_{j[J}^{kL} e^0_{I]} \psi_L^K e^j_K \,\right)
\ee
and:
\be
{\left(\mathbb{X}^{\textsc{d}}_\xi\right)_a}^{IJ}_k \,=\, \xi^j F_{kj}^{IJ} \,, \qquad {\left(\mathbb{X}^{\textsc{d}}_\xi\right)_p}^k_{IJ} \,=\, -\left(\, U^K_{[I} e^k_{J]} \xi^j \nabla_j e^0_K - W^{kK}_{j[J} e^0_{I]} \xi^l \nabla_l e^j_K \,\right) \,,
\ee
where $U_{[I}^L \,=\, \delta_{[I}^L - e^0_{[I} e^L_0$, $W_{j[J}^{kL} \,=\, \delta_j^k \delta_{[J}^L + e_j^L e^k_{[J}$ and being actually tangent to $\mathcal{M}_\infty$.
The latter tangent vectors span, at each point, ${\mathbf{T}_{\overline{\chi}_\Sigma} \mathcal{M}_\infty^\perp}^\parallel$ and, therefore, the whole $\mathbf{T}_{\overline{\chi}_\Sigma} \mathcal{M}_\infty^\perp$ is now characterized.

To resume the above discussion, we saw that solutions of the pre-symplectic Hamiltonian system $\left(\mathcal{M},\, {\os}^{\mathrm{ext}},\, \mathcal{H}^{\mathrm{ext}}\right)$, which, at least locally, coincide with extrema of the tetradic Palatini's action, are integral curves of the vector field \eqref{Eq:transversal tangent vector} restricted to the final manifold $\mathcal{M}_\infty$ of the PCA, that is, torsionless solutions of Einstein's equations (see \eqref{Eq:torsionless + einstein}), up to gauge transformations \eqref{Eq:gauge palatini} and \eqref{Eq:diffeo palatini} obtained as the tangent part of the kernel of the pre-symplectic form.

\subsection{Covariant brackets and DeWitt's formula}
\label{Subsec:Covariant brackets and DeWitt's formula}

With the pre-symplectic structure ${\os}^{\mathrm{ext}}_\infty$ in hand, we can use the machinery developed in \cite{Ciaglia-DC-Ibort-Marmo-Schiav-Zamp2021-Cov_brackets_toappear, Ciaglia-DC-Ibort-Mar-Schiav-Zamp2022-Non_abelian} to equip the manifold $\mathcal{M}_\infty$, i.e. what we referred to as the solution space of the theory in \cite{Ciaglia-DC-Ibort-Marmo-Schiav-Zamp2021-Cov_brackets_toappear, Ciaglia-DC-Ibort-Mar-Schiav-Zamp2022-Non_abelian}, with a Poisson structure.
In particular, we recall that given a connection $P$ on the bundle $\mathcal{M}_\infty \to \mathcal{M}_\infty / \mathrm{ker}{\os}^{\mathrm{ext}}_\infty$, it is possible to construct a Poisson structure on an enlargement of $\mathcal{M}_\infty$ obtained out of the coisotropic embedding theorem.
Such a Poisson structure projects to a Poisson structure on $\mathcal{M}_\infty$ if the connection is flat.
When this is the case we saw that the Poisson bracket on $\mathcal{M}_\infty$ is given by the expression \eqref{Eq:covariant bracket} together with the condition horizontality condition with respect to $P$ \eqref{Eq:horizontality}.

Without referring to a particular choice of $P$ that we still do not have in hand, we end this section by conjecturing that a flat connection $P$ can be chosen on the bundle mentioned above and that DeWitt's condition \eqref{Eq:DeWitt condition} can be written as the horizontality condition for such a connection.

\section*{Conclusions}
\label{Sec:Conclusions}
\addcontentsline{toc}{section}{\nameref{Sec:Conclusions}}

This manuscript represented the last contribution of a series of papers where the authors dealt with the problem of equipping the space of solutions of the equations of motion of a large class of classical field theories with a Poisson structure.
In particular, having in mind the relation with quantum field theories (see the Introduction of \cite{Ciaglia-DC-Ibort-Marmo-Schiav-Zamp2021-Cov_brackets_toappear}), the authors were interested in dealing with those field theories associated with the description of fundamental interactions.
In this respect, this last manuscript cover the gap of the first two contributions where Gravitational interaction was not considered.

Indeed, in this paper we showed how, taking advantage of Palatini's formulation of General Relativity, Gravitation can be formulated as a particular type of constrained Yang-Mills theory and, consequently, how to adapt the whole machinery developed in \cite{Ciaglia-DC-Ibort-Marmo-Schiav-Zamp2021-Cov_brackets_toappear, Ciaglia-DC-Ibort-Mar-Schiav-Zamp2022-Non_abelian} to such a "constrained formalism".
This allowed us to provide a way to equip the space of solutions of Einstein's equations in vacuum with a Poisson structure and to conjecture a way to geometrically interpret the bracket written by \textit{B. DeWitt} in \cite{DeWitt1960-Commutators_Quant_Grav}.

\bibliographystyle{alpha}
\bibliography{Biblio}

\newcommand{\etalchar}[1]{$^{#1}$}
\begin{thebibliography}{CDI{\etalchar{+}}20b}

\bibitem[BM94]{Baez1994-Gauge_theories}
J.~Baez and J.~P. Muniain.
\newblock {\em {Gauge Fields, Knots and Gravity}}.
\newblock Series on Knots and Everything. World Scientific, Singapore, 1994.

\bibitem[CDI{\etalchar{+}}a]{Ciaglia-DC-Ibort-Marmo-Schiav-Zamp2021-Cov_brackets_toappear}
F.~M. Ciaglia, F.~{Di Cosmo}, L.~A. Ibort, G.~Marmo, L.~Schiavone, and
  A.~Zampini.
\newblock {The geometry of the solution space of first order Hamiltonian field
  theories I: from particle dynamics to Electrodynamics}.
\newblock {\em arXiv:2208.14136}.

\bibitem[CDI{\etalchar{+}}b]{Ciaglia-DC-Ibort-Mar-Schiav-Zamp2022-Non_abelian}
F.~M. Ciaglia, F.~{Di Cosmo}, L.~A. Ibort, G.~Marmo, L.~Schiavone, and
  A.~Zampini.
\newblock {The geometry of the solution space of first order Hamiltonian field
  theories II: non-Abelian gauge theories}.
\newblock {\em arXiv:2208.14155}.

\bibitem[CDI{\etalchar{+}}20a]{Ciaglia-DC-Ibort-Marmo-Schiav2020-Jacobi_Fields}
F.~M. Ciaglia, F.~{Di Cosmo}, L.~A. Ibort, G.~Marmo, and L.~Schiavone.
\newblock {Covariant Variational Evolution and Jacobi brackets: Fields}.
\newblock {\em Modern Physics Letters A}, 35(23):1 --16 --
  10.1142/S0217732320502065, 2020.

\bibitem[CDI{\etalchar{+}}20b]{Ciaglia-DC-Ibort-Marmo-Schiav2020-Jacobi_Particles}
F.~M. Ciaglia, F.~{Di Cosmo}, L.~A. Ibort, G.~Marmo, and L.~Schiavone.
\newblock {Covariant variational evolution and Jacobi brackets: Particles}.
\newblock {\em Modern Physics Letters A}, 35(23):1 --17 --
  10.1142/S0217732320200011, 2020.

\bibitem[CDI{\etalchar{+}}22]{Ciaglia-DC-Ibort-Marmo-Schiav-Zamp-2022-Symmetry}
F.~M. Ciaglia, F.~{Di Cosmo}, L.~A. Ibort, G.~Marmo, L.~Schiavone, and
  A.~Zampini.
\newblock {Symmetries and Covariant Poisson Brackets on Presymplectic
  Manifolds}.
\newblock {\em Symmetry}, 14(70):1 -- 28, 2022.

\bibitem[DeW60]{DeWitt1960-Commutators_Quant_Grav}
B.~S. DeWitt.
\newblock {Invariant commutators for the Quantized Gravitational Field}.
\newblock {\em Physical Review Letters}, 4(6):317--320, 1960.

\bibitem[DeW65]{DeWitt1965-Groups_Fields}
B.~S. DeWitt.
\newblock {\em {Dynamical theory of groups and fields}}.
\newblock Gordon and Breach, New York, 1965.

\bibitem[GMS10]{Giac-Mang-Sard2010-Geometric_Classical_Quantum_Mechanics}
G.~Giachetta, L.~Mangiarotti, and G.~Sardanashvily.
\newblock {\em {Geometric formulation of Classical and Quantum Mechanics}}.
\newblock World Scientific Publishing Co., Singapore, 2010.

\bibitem[Kru15]{Krupka2015-Variational_Geometry}
D.~Krupka.
\newblock {\em {Introduction to global variational geometry}}.
\newblock Atlantis Press, 2015.

\bibitem[Pei52]{Peierls1952-Commutation_laws}
R.~E. Peierls.
\newblock {The commutation laws of relativistic field theory}.
\newblock {\em Proceedings of the Royal Society of London. Series A.
  Mathematical and Physical Sciences}, 214(1117):143--157, 1952.

\bibitem[Rov04]{Rovelli2004-Quantum_Gravity}
C.~Rovelli.
\newblock {\em {Quantum Gravity}}.
\newblock Cambridge University Press, Cambridge, 2004.

\bibitem[Spi99]{Spivak1999-Vol2}
M.~Spivak.
\newblock {\em {A comprensive introduction to Differential Geometry}}.
\newblock Publish or Perish, Inc., Houston, 1999.

\end{thebibliography}

\end{document}